\documentclass[10pt,conference]{IEEEtran}
\IEEEoverridecommandlockouts

\usepackage{cite}
\usepackage{amsmath,amssymb,amsfonts}
\usepackage{algorithmic}
\usepackage{graphicx}
\usepackage{textcomp}
\usepackage{xcolor}
\usepackage{multirow}
\usepackage{tcolorbox}
\usepackage{listings}
\usepackage{color}
\usepackage{enumitem}
\usepackage{seqsplit}
\usepackage{subcaption}
\usepackage{float}
\usepackage[table]{xcolor}
\usepackage{caption}
\usepackage{booktabs}
\usepackage{url}
\usepackage{amsmath}

\definecolor{customcolor}{RGB}{106,122,191}
\definecolor{forestgreen}{RGB}{69, 135, 52}
\definecolor{newyellow}{RGB}{243, 167, 60}

\newtcolorbox{promptbox}[1]{colframe=customcolor,fonttitle=\bfseries\large,title={#1}}

\newcommand{\best}[1]{\textbf{#1}}
\newcommand\etal{{\it{et al.\ }}}
\newcommand{\LCM}[1]{LCM\ifx#1ss\fi}

\begin{document}

\title{\fontsize{23}{30}\selectfont Automated Prompt Generation for Code Intelligence: An Empirical study and Experience in WeChat
\thanks{$^{*}$Cuiyun Gao is the corresponding author.}
}



\author{
Kexing Ji$^1$, Shiyun Fu$^1$, Cuiyun Gao$^{1*}$, Yujia Chen$^1$, Zezhou Yang$^2$, Chaozheng Wang$^2$, Yuetang Deng$^2$ \\
$^1$The Chinese University of Hong Kong, Hong Kong, China \\
$^2$WeChat Group, Tencent Inc., Guangzhou, China \\
kexing1208@gmail.com, FuShiyun1@outlook.com, cuiyungao@outlook.com, yujiachenhit@gmail.com, \\ zezhouyang@tencent.com, adf111178@gmail.com, yuetangdeng@tencent.com
}




\maketitle


\begin{abstract}
Large Code Models (\LCM{s}) have demonstrated potential in advancing various code intelligence tasks. However, their effectiveness can be greatly influenced by the quality of the prompts. Current prompt design strategies in code intelligence studies are mostly manually generated, which could be time-consuming and extremely rely on the base \LCM{s} and tasks. Although automated prompt generation (APG) has been investigated in the natural language processing field, it has not attracted sufficient attention and been well explored in the code intelligence tasks. Considering the various tasks and black-box nature of LCMs faced by developers in practice, it is essential to automate the prompt generation process.

To mitigate the gap, we empirically investigate the two important parts in APG, including Instruction Generation (IG) and Muti-Step Reasoning (MSR). The instruction generation part aims at providing a task-related description for instructing \LCM{s} to effectively accomplish specific tasks; while the multi-step reasoning part aims at guiding \LCM{s} to produce a series of logical steps before arriving at the final answer. For each part, we evaluate the widely-used APG methods on four open-source \LCM{s} and three code intelligence tasks, i.e., code translation (PL-PL), code summarization (PL-NL) and API recommendation (NL-PL).
Experimental results indicate that the two parts in APG can dramatically enhance the performance of the code intelligence tasks compared with the basic prompts. 
Based on the results, we further propose a novel APG approach by combining the best methods of the two studied parts of APG.
Experiments show that the proposed APG approach achieves an average improvement of 28.38\% with respect to CodeBLEU for the code translation, 58.11\% in terms of ROUGE-L for the code summarization and 84.53\% in SuccessRate@1 for the API recommendation over the basic prompts, respectively.
To validate the effectiveness in industrial scenario, we further evaluate our approach on WeChat-Bench, a proprietary dataset from the WeChat Group in Tencent for API recommendation, achieving an average improvement of 148.89\% in MRR.

\end{abstract}


\begin{sloppypar}
\section{Introduction}

\begin{figure}
    \centering
    \includegraphics[width=0.5\textwidth]{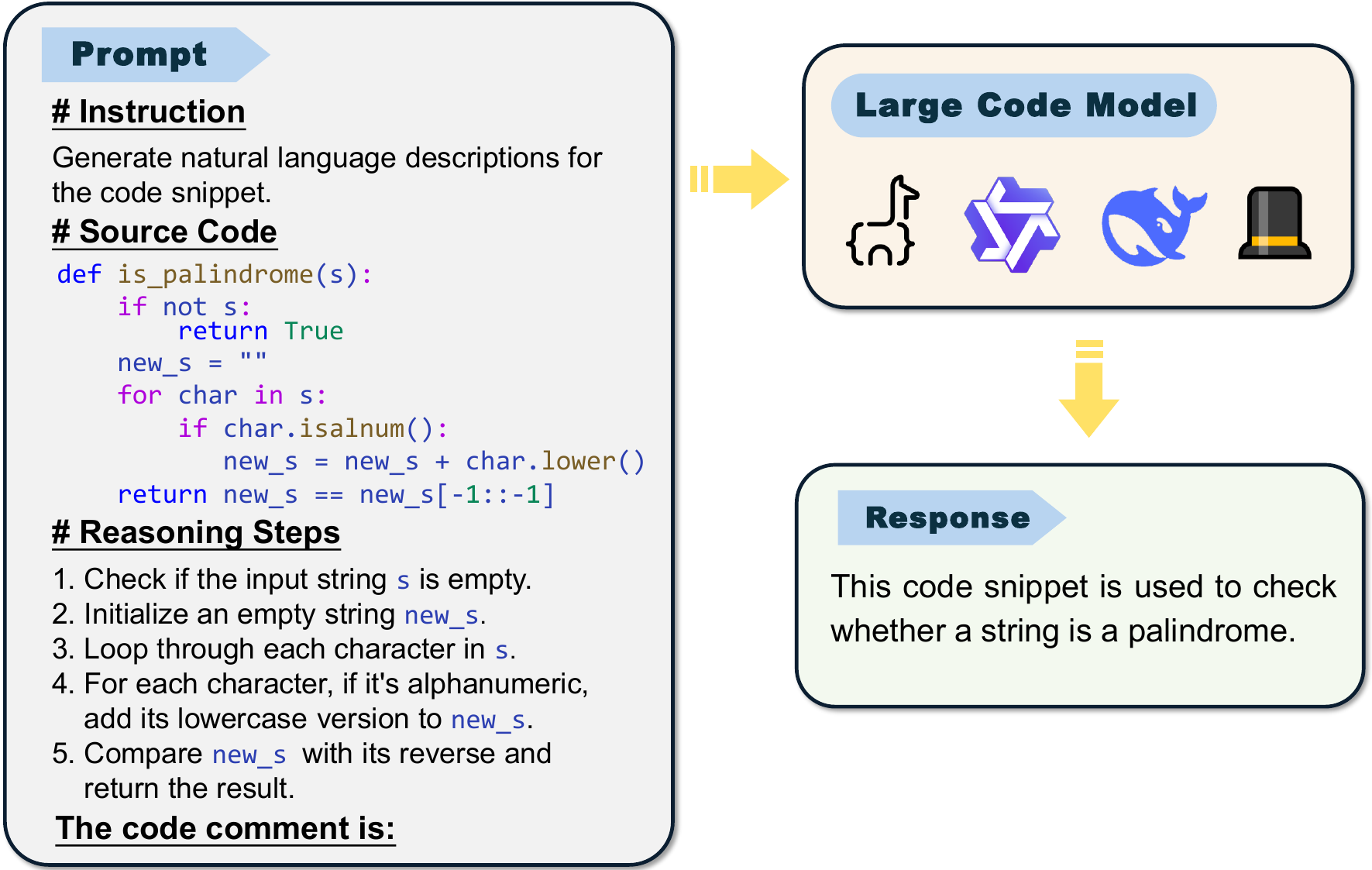}
    \caption{An example of designed prompt for \LCM{} to generate the summary of given code snippet.}
    \label{fig:prompt}
\end{figure}

Recently, the advent of large-scale code corpora combined with advances in deep learning technologies has facilitated the development of Large Code Models (\LCM{s}), e.g., Code Llama~\cite{DBLP:journals/corr/abs-2308-12950}, DeepSeek-Coder~\cite{DBLP:journals/corr/abs-2401-14196}, and Qwen2.5-Coder~\cite{DBLP:journals/corr/abs-2409-12186}.
These \LCM{s} demonstrate great potential in advancing various code intelligence tasks, including code summarization~\cite{DBLP:conf/acl/AhmadCRC20, DBLP:journals/corr/abs-2308-13775}, API recommendation~\cite{DBLP:conf/emnlp/KangW00Y21, DBLP:journals/tosem/LiLTHGLNWHZ24} and code translation~\cite{DBLP:conf/nips/RoziereLCL20, DBLP:conf/emnlp/YangHHHJ20}. 
To effectively employ LCMs in these tasks, users generally provide task-specific prompts that include both instructions and contextual information. 
Figure~\ref{fig:prompt} provides an example of this interactive process. The user formulates a prompt containing the task description, source code, and reasoning steps for the code summarization task. Subsequently, the LCM receives the prompt as input and generates a response.

Current prompt design strategies~\cite{DBLP:conf/sigsoft/WangYGP0L22,DBLP:conf/icse/PanIKSWMSPSJ24,DBLP:conf/icse/GengWD00JML24} in code intelligence are generally manually written.
For instance, Sun \etal~\cite{DBLP:journals/corr/abs-2305-12865} manually design 11 different prompts and conduct experiments to decide which prompt is more suitable for specific \LCM{s} to generate code comments.
White~\etal~\cite{DBLP:journals/corr/abs-2303-07839} manually craft 13 types of prompts for four task categories, including code quality improvement, refactoring, requirements acquisition, and software design.
Although the human-written prompts achieve superior performance in code intelligence tasks, they bring several challenges.
First, different human-written prompts can lead to dramatically different performance. 
For example, Sun~\etal~\cite{DBLP:journals/corr/abs-2305-12865} report that the least effective prompt results in a 119.87\% reduction in BLEU scores compared to the most effective prompt.
Second, in industrial development scenarios, different projects often require task-specific and model-specific prompts tailored to their unique codebases, APIs, and development requirements, making manually-generated prompts hard to be reused across projects~\cite{DBLP:journals/corr/abs-2304-11938}. 
Moreover, the manual process is time-consuming and may bring much labor cost, particularly when scaling across various code intelligence tasks and \LCM{s}.
Therefore, how to automatically generate effective prompts for different tasks and \LCM{s} becomes an important research question.


Recent studies in the natural language processing (NLP) field have proposed various Automated Prompting Generation (APG) methods~\cite{DBLP:journals/corr/abs-2303-09014, DBLP:conf/iclr/ZhouMHPPCB23, DBLP:conf/iclr/YaoZYDSN023, DBLP:conf/acl/0010LLWWBCH22, DBLP:conf/emnlp/ShinRLWS20} and have achieved promising results in NLP tasks.
The APG methods leverage the task definitions and input data to automatically generate prompts for LCMs, enabling adaptability across different models and tasks.
However, while APG methods have proven effective in the NLP field, the application of these methods to code intelligent tasks is still unexplored.
Considering the importance of prompt design in code intelligence tasks, it is necessary to explore the APG methods for code intelligence tasks.

To this end, we conduct a comprehensive empirical study on the impact of APG for code intelligence tasks and evaluate its effectiveness in industrial scenario.
Based on the constituent parts of a prompt, i.e., instruction and reasoning steps as illustrated in Fig.~\ref{fig:prompt}, we focus on two important parts in APG, including Instruction Generation (IG) and Multi-Step Reasoning (MSR) in this paper. 
Specifically, instruction generation aims at providing a task-related description for instructing \LCM{s} to effectively accomplish specific tasks, while multi-step reasoning aims at guiding \LCM{s} to produce a series of intermediate logical steps before arriving at the final answer.

For each aspect, we evaluate the widely-used APG methods on four open-source \LCM{s}.
We choose three popular code intelligence tasks, including API recommendation (NL-PL), code translation (PL-PL), and code summarization (PL-NL) for evaluation. We focus on investigating the following four research questions (RQs):

\begin{enumerate}
    \item What is the impact of instruction generation on the effectiveness of APG?
    \item How effective is multi-step reasoning in enhancing the performance of APG?
    \item Can we further improve the performance of current APG methods?
    \item How does the improved APG method perform in the industrial scenario?
\end{enumerate}

To explore the impact of instruction generation on the effectiveness of APG methods (RQ1), we evaluate two representative automated instruction generation methods: Automatic Prompt Engineer~\cite{DBLP:conf/iclr/ZhouMHPPCB23} and Optimization by PROmpt (OPRO)~\cite{DBLP:journals/corr/abs-2309-03409}, and compare them against manually crafted prompts~\cite{DBLP:conf/icse/PanIKSWMSPSJ24, DBLP:conf/kbse/GaoWGWZL23} across four open-source LCMs.
To investigate the effectiveness of multi-step reasoning in enhancing the performance of LCMs (RQ2), we evaluate the three widely adopted multi-step reasoning methods—Chain-of-Thought (CoT)~\cite{DBLP:conf/nips/Wei0SBIXCLZ22}, AutoCoT~\cite{DBLP:conf/iclr/0001Z0S23}, and Self-Plan~\cite{DBLP:journals/corr/abs-2303-06689} and compare their performance with that of manually designed prompts.
To further explore whether the performance of current APG methods can be improved (RQ3), 
we combine the best-performing instruction generation and multi-step reasoning methods identified in RQ1 and RQ2, and evaluate them on four open-source LCMs for code intelligence tasks.
Building upon the findings of RQ3, we further investigate whether the observed improvements in APG methods can generalize to the industrial scenario (RQ4), thereby evaluating the practical applicability of the proposed technique.

Based on the experimental results, we find that APG methods greatly improve LCM performance across three code intelligence tasks.
Among the evaluated APG methods, APE proves to be the most effective instruction generation method, while CoT achieves the best results among multi-step reasoning methods.
Furthermore, the novel approach APE-CoT achieve non-trivial performance improvements over basic prompts, with an average increase of 28.38\% in CodeBLEU for code translation, 58.11\% in ROUGE-L for code summarization and 84.53\% in SucessRate@1 for API recommendation.
Additionally, evaluation on WeChat-Bench, a proprietary dataset of 1000 C++ samples from the WeChat Group in Tencent for API recommendation, demonstrates an average improvement of 148.89\% in MRR, confirming the effectiveness of our approach in the industrial scenario.

The main contributions of our work are summarized below:

\begin{enumerate}
    \item To the best of our knowledge, this paper presents the first in-depth investigation into 
    how to effectively utilize APG to generate prompts for code intelligence tasks.
    \item Based on our findings, we propose a novel APG method that
    greatly improves the performance of \LCM{s} in code intelligence tasks compared to the basic prompts.
    \item We validate our approach in industrial scenario through evaluation in WeChat and offer practical implications for developers and potential directions for researchers.

\end{enumerate}

\section{Background}\label{sec:background}
\begin{figure*}
    \centering
    \includegraphics[width=0.98\textwidth]{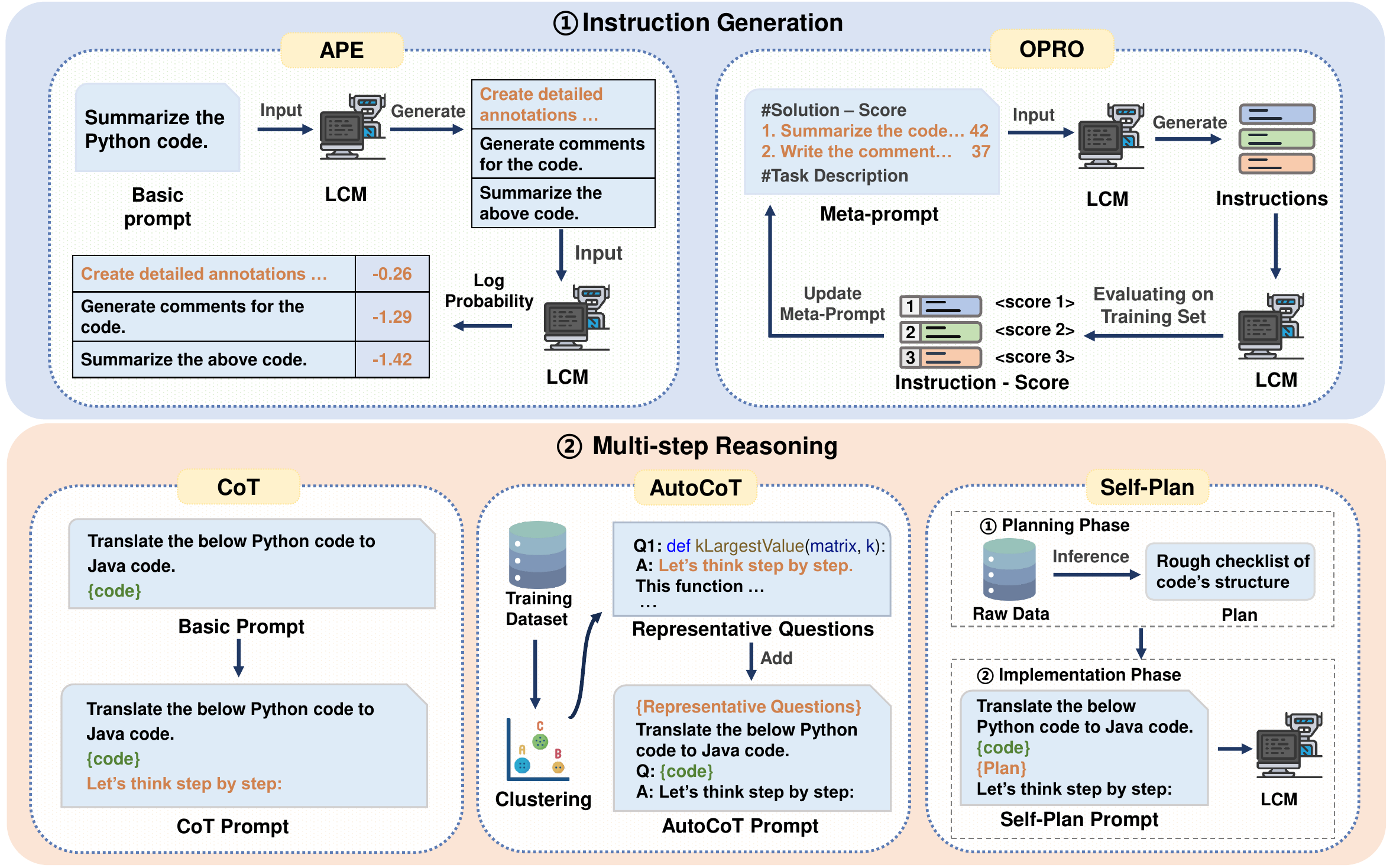}
    \caption{The overview of APG Methods: Instruction Generation and Multi-step Reasoning.}
    \label{fig:overview}
\end{figure*}

As shown in Fig.~\ref{fig:prompt}, APG contains two key parts: \textit{instruction generation}, which creates a tailored task description based on the input, and \textit{multi-step reasoning}, which produces structured steps to solve the task. We elaborate on the details of the two parts in the following.


\subsection{Instruction Generation}
For instruction generation, we consider two widely acknowledged and effective techniques, namely APE~\cite{DBLP:conf/iclr/ZhouMHPPCB23} and OPRO~\cite{DBLP:journals/corr/abs-2309-03409}, as detailed below.

\begin{itemize}[leftmargin=*]
    \item  \textbf{Automatic Prompt Engineer (APE)~\cite{DBLP:conf/iclr/ZhouMHPPCB23}} treats instruction generation as a natural language synthesis task.
    As shown in Figure~\ref{fig:overview}, 
    \LCM{s} first automatically generate a pool of candidate instructions through comprehending the task requirements underneath basic prompt, and then select the best instruction through log probability. 
    The log probability reflects the likelihood of producing the desired result given the instruction and the LCM.
    \item  \textbf{Optimization by PROmpting (OPRO)~\cite{DBLP:journals/corr/abs-2309-03409}} 
    uses LCMs to refine task execution through dynamic optimization. 
    Figure 2 illustrates the OPRO process. A meta-prompt guides LCMs to generate and evaluate candidate instructions, and the top-performing pairs are used to update the meta-prompt iteratively until convergence on an optimal instruction.
    
\end{itemize}

\subsection{Multi-step Reasoning}
For multi-step reasoning, we select three representative methods—CoT, AutoCoT, and Self-Plan—based on their remarkable performance in prior work~\cite{DBLP:conf/icse/LiW0L0WG024}, as detailed below. 

\begin{itemize}[leftmargin=*]
    \item \textbf{Chain-of-Thought (CoT)}~\cite{DBLP:conf/nips/Wei0SBIXCLZ22} 
    enhances the reasoning capabilities of LCMs by guiding them through a sequence of intermediate steps, thereby reducing the difficulty of generating the desired output.
    For example, as shown in Figure~\ref{fig:overview}, by simply adding ``Let's think step by step'', CoT prompt encourages \LCM{s} to produce more considered responses by focusing on intermediary reasoning steps.
    \item \textbf{AutoCoT~\cite{DBLP:conf/iclr/0001Z0S23}}
    serves as an automated multi-step reasoning method that optimizes CoT in terms of efficiency. As
    shown in Figure~\ref{fig:overview}, AutoCoT clusters the given training dataset and selects representative problems from each cluster. These representative questions, along with their reasoning steps, are then incorporated into the AutoCoT prompt to better guide \LCM{s} in generating accurate outputs.
    \item \textbf{Self-Plan~\cite{DBLP:journals/corr/abs-2303-06689}} 
    decomposes complex tasks into a structured subtasks, outlining the overall program structure without delving into implementation details. It provides clear and concise high-level actions through a two-stage generation process, as illustrated in Figure 2. In the first stage, the LCM generates a “plan” that serves as an outline for the code structure. In the second stage, the LCM leverages both the plan and code to generate outputs for downstream tasks.
    
\end{itemize}

\section{EXPERIMENTAL SETUP}\label{sec:setup}
\subsection{Research Questions}
In this paper, we conduct
experiments
with the aim of answering the following research questions:

\begin{itemize}
\item \textbf{RQ1:} What is the impact of instruction generation on the effectiveness of APG?
\item \textbf{RQ2:} How effective is multi-step reasoning in enhancing the performance of APG?
\item \textbf{RQ3:} Can we further improve the performance of current APG methods?
\item \textbf{RQ4:} How does the improved APG method perform in an  industrial scenario?
\end{itemize}

\subsection{Evaluation Tasks}
\label{sec:tasks}

We evaluate the APG approaches on three representative code intelligence tasks: code summarization, code translation and API recommendation.

%
\subsubsection{Code Summarization}
Code summarization, also known as code comment generation, targets to automatically generate concise natural language descriptions for source code snippets~\cite{DBLP:journals/tosem/GaoGHZNXL23,geng2023empirical}. 
We use the prompt template 
from~\cite{DBLP:conf/sigsoft/WangYGP0L22} as our basic prompt: 

\begin{center}
    ``\textit{Generate comments for} [LANG] \textit{code.}''
\end{center}
where [LANG] denotes the slot of programming language.


\textbf{Datasets.} 
We use the CodeXGLUE code summarization benchmark~\cite{DBLP:journals/corr/abs-2102-04664}. It contains code snippets and corresponding natural language comments extracted from GitHub repositories.
The detailed data statistics are shown in Table~\ref{tab:dataset}.

\textbf{Metrics.} 
Following previous work~\cite{DBLP:conf/kbse/GaoWGWZL23}, we use three widely adopted metrics for evaluating code summarization: BLEU-4~\cite{DBLP:conf/acl/PapineniRWZ02}, ROUGE-L~\cite{lin-2004-rouge}, and METEOR~\cite{DBLP:conf/acl/BanerjeeL05}. These metrics evaluate the similarity between generated and ground-truth summaries and are widely used in code summarization.


\begin{table}[t]
\centering
\caption{Statistics of the datasets.}
\resizebox{\columnwidth}{!}{%
\begin{tabular}{c|c|rrr}
\toprule
Task           & Dataset                 & Train                   & Valid                & Test                   \\ \midrule
Code           & CodeXGLUE-Java          & 164,923                 & 5,183                & 10,955                 \\
Summarization  & CodeXGLUE-Python        & 251,820                 & 13,914               & 14,918                 \\ \midrule
Code           & AVATAR-Java             & 60,138                  & 476                  & 1,906 
                \\ 
Translation    & AVATAR-Python           & 60,138                  & 476                  & 1,906                    \\ \midrule
API            & APIBENCH-Q-Java         & 4,769                   & 594                    & 600                    \\
Recommendation & APIBENCH-Q-Python       & 3,118                   & 391                    & 400                    \\ 
\bottomrule
\end{tabular}%
}
\label{tab:dataset}
\end{table}


\subsubsection{Code Translation}
Code translation aims to convert code written in one programming language to another~\cite{DBLP:conf/iui/WeiszMHRRMAT21}. This involves interpreting the logic and functionality of the original code and reproducing it accurately in the target language. We utilize the template described in~\cite{DBLP:conf/icse/PanIKSWMSPSJ24} as the basic prompt:
\begin{center}
    ``\textit{Translate the above} [SOURCE] \textit{code to} [TARGET].''
\end{center}
where [SOURCE] and [TARGET] specify the source and target programming languages for translation, respectively.

\textbf{Datasets.} The dataset for the code translation task is Avatar~\cite{DBLP:conf/acl/AhmadTCC23}. 
We use parallel functions in Java and Python to evaluate the task.
The detailed data statistics are in shown Table~\ref{tab:dataset}.

\textbf{Metrics.} To evaluate the code translation task, we adopt three additional metrics besides BLEU: Syntax Match (SM), Dataflow Match (DM), and CodeBLEU (CB)~\cite{DBLP:journals/corr/abs-2009-10297}. SM measures the proportion of matching subtrees in Abstract Syntax Trees (AST), while DM calculates the proportion of matching data flow edges. CB extends BLEU by incorporating syntactic and semantic features such as AST and data flow.

\subsubsection{API Recommendation}
API recommendation aims to suggest appropriate Application Programming Interface (API) calls for specific programming tasks. In this work, we focus on natural-language-based API recommendation, where the goal is to retrieve suitable APIs according to user' textual queries. We use the template described in ~\cite{DBLP:conf/wcre/ChenGZ0WX24} as the basic prompt.

\begin{center}
    ``\textit{Please recommend some suitable APIs for the given query.}''
\end{center}

\textbf{Datasets.} In API recommendation task, we adopt the widely used APIBENCH-Q~\cite{DBLP:journals/tse/PengLGLWGL23} dataset. 
Following an 8:1:1 train/validation/test split, we randomly select 400 Python and 600 Java questions as the test set, with the rest queries grouped into the training and validation set.
The detailed data statistics are shown in Table~\ref{tab:dataset}.

\textbf{Metrics.} 
Following prior work~\cite{DBLP:conf/wcre/ChenGZ0WX24}, we utilize the widely used metrics in recommendation tasks to evaluate~\cite{DBLP:journals/tse/PengLGLWGL23,DBLP:conf/sigsoft/GuZZK16,DBLP:conf/icse/WeiHH0022}: Success Rate (SR) and Mean Reciprocal Rank (MRR). 
SR@k evaluates a method's ability to suggest relevant APIs within the top-k recommendations, with k values of 1, 3, and 5 used for our evaluation.
Meanwhile, MRR computes the average reciprocal rank of the first relevant API across all queries.

\subsection{Large Code Models}
To evaluate the effectiveness of APG methods, we utilize four open-source LCMs: 


\begin{itemize}[leftmargin=*]

\item \textbf{CodeLlama}~\cite{DBLP:journals/corr/abs-2308-12950}
is a family of large language models for code based on Llama 2~\cite{DBLP:journals/corr/abs-2307-09288} with state-of-the-art code generation, and blank infilling capabilities.

\item \textbf{Qwen2.5-Coder}~\cite{DBLP:journals/corr/abs-2409-12186}
is a LCM 
from Qwen2.5 series with 128K context length. 
It is trained on 5.5T tokens over source code and text-code alignment data, achieving strong performance in code generation and reasoning.

\item 
\textbf{Deepseek-Coder}~\cite{DBLP:journals/corr/abs-2401-14196}
is a series of LCMs trained on 2T tokens across over 80 programming languages. It achieves state-of-the-art performance among open-source code LLMs.

\item \textbf{Magicoder} \cite{DBLP:journals/corr/abs-2312-02120}
is an instruction-tuned LCM, which first trains Deepseek-Coder on 75K OSS-INSTRUCT data and further trains the model on 110K Evol-Instruct data.





\end{itemize}

\subsection{Implementation Details}

In our study, We employ the following four open-source \LCM{s}: Deepseek-Coder (Deepseek-Coder-6.7B-Instruct), Magicoder (\seqsplit{Magicoder-S-DS-6.7B}), CodeLlama (CodeLlama-13b-Instruct) and Qwen2.5-Coder (Qwen2.5-Coder-14B-Instruct). We download these models from HuggingFace\footnote{https://huggingface.co/models} and deploy them locally. 
For the implementation of instruction generation (APE and OPRO) and multi-step reasoning methods (CoT, AutoCoT and Self-Plan), we directly use the replication packages released by the authors and adapt them to our tasks.
As for the hyperparameters for the generation of \LCM{s}, we adopt nuclear sampling with a top-p value of 0.95 and a temperature value of 0.2. The maximum generation token length is set to 256 for code translation, 512 for code summarization, and 128 for API recommendation.

All the experiments are conducted on an Ubuntu-20.04 server equipped with 4 * Nvidia A100 GPUs and each one has 40GB graphics memory. The source code is publicly accessible at \url{https://anonymous.4open.science/r/Towards-Prompt-is-All-You-Need-5D66}.

\begin{table*}[t]
    \centering
    \caption{Experiment results of instruction generation methods in code intelligence tasks.}
    \label{tab:IG}
    \resizebox{0.9\textwidth}{!}{
        \begin{tabular}{c|c|cccc|cccc|ccc}
            \toprule
            
            \rule[-6pt]{0pt}{10pt}
            
            \multirow{3}{*}{\textbf{Model}} & \multirow{3}{*}{\textbf{Approach}} & \multicolumn{4}{c|}{\textbf{API Recommendation (NL-PL)}} & \multicolumn{4}{c|}{\textbf{Code Translation (PL-PL)}} & \multicolumn{3}{c}{\textbf{Code Summarization (PL-NL)}} \\
            \cline{3-13}
            
             & & \multicolumn{4}{c|}{\textbf{Python}} & \multicolumn{4}{c|}{\textbf{Python $\rightarrow$ Java}} & \multicolumn{3}{c}{\textbf{Python}} \\
            & & SR@1 & SR@3 & SR@5 & MRR & CB & SM & DM & BLEU & BLEU & ROUGE-L & METEOR \\
            \midrule
            
            \multirow{3}{*}{Deepseek-Coder} & Basic &
            12.75 & 15.50 & 17.00 & 14.38 &
            51.43 & 74.29 & 50.79 & 39.45 &
            15.68 & 18.82 & 7.06 \\
            
            & APE & 
            \best{15.25} & \best{17.75} & \best{18.75} & \best{16.56} &
            \best{65.36} & \best{76.05} & \best{68.00} & \best{58.92} &
            \best{15.87} & \best{20.60} & \best{8.66} \\
            
            & OPRO & 
            13.50 & 18.00 & 18.75 & 15.69 &
            60.19 & 75.42 & 60.86 & 51.44 &
            15.85 & 20.30 & 8.24 \\
            \midrule
            
            \multirow{3}{*}{Magicoder} & Basic & 
            11.75 & 14.25 & 16.00 & 13.22 &
            52.51 & 68.85 & 33.10 & 48.58 &
            14.77 & 18.56 & 7.09 \\
            
            & APE & 
            \best{14.50} & \best{17.00} & \best{19.75} & \best{16.23} &
            \best{63.02} & \best{76.57} & \best{68.35} & \best{52.98} &
            \best{15.83} & \best{20.44} & \best{8.54} \\
            
            & OPRO & 
            14.25 & 16.00 & 17.75 & 15.36 &
            60.51 & 76.04 & 61.35 & 51.75 &
            15.78 & 19.75 & 7.73 \\
            \midrule
            
            \multirow{3}{*}{CodeLlama} & Basic & 
            10.50 & 13.75 & 14.75 & 12.00 &
            56.29 & 70.05 & 57.48 & 48.20 &
            9.24 & 10.78 & 4.75 \\
            
            & APE & 
            \best{17.50} & \best{19.50} & \best{22.00} & \best{18.99} &
            \best{62.94} & \best{74.15} & \best{65.08} & \best{56.27} &
            \best{15.82} & \best{20.44} & \best{7.52} \\
            
            & OPRO &
            15.25 & 17.25 & 18.25 & 16.39 &
            58.60 & 72.24 & 61.61 & 50.00 &
            15.82 & 19.96 & 6.56 \\
            \midrule
            
            \multirow{3}{*}{Qwen2.5-Coder} & Basic &
            10.25  & 13.25 & 14.50 & 12.31 &
            56.18 & 70.29 & 57.52 & 48.73 &
            12.47 & 14.36 & 5.82 \\
            
            & APE & 
            \best{18.75} & \best{20.50} & \best{21.75} & \best{19.12} &
            \best{64.95} & \best{75.16} & \best{66.91} & \best{56.48} &
            \best{15.91} & \best{20.18} & \best{8.63} \\
            
            & OPRO & 
            15.50 & 17.25 & 18.50 & 16.08 &
            59.44 & 72.68 & 63.84 & 50.36 &
            14.76 & 18.68 & 7.58 \\
            \midrule
            
            \rule[-6pt]{0pt}{10pt}
            
            \multirow{3}{*}{\textbf{Model}} & \multirow{3}{*}{\textbf{Approach}} & \multicolumn{4}{c|}{\textbf{API Recommendation (NL-PL)}} & \multicolumn{4}{c|}{\textbf{Code Translation (PL-PL)}} & \multicolumn{3}{c}{\textbf{Code Summarization (PL-NL)}}\\
            \cline{3-13}

            & & \multicolumn{4}{c|}{\textbf{Java}} & \multicolumn{4}{c|}{\textbf{Java $\rightarrow$ Python}} & \multicolumn{3}{c}{\textbf{Java}} \\
            & & SR@1 & SR@3 & SR@5 & MRR & CB & SM & DM & BLEU & BLEU & ROUGE-L & METEOR \\
            \midrule
            
            \multirow{3}{*}{Deepseek-Coder} & Basic & 
            19.67 & 22.83 & 24.33 & 21.39 &
            49.88 & 59.39 & 37.11 & 50.75 &
            14.15 & 15.57 & 6.12 \\
            
            & APE & 
            \best{21.67} & \best{26.50} & \best{28.00} & \best{24.22} &
            \best{58.61} & \best{61.98} & \best{41.94} & \best{64.92} &
            \best{15.02} & 18.60 & \best{7.24} \\
            
            & OPRO & 
            20.17 & 23.50 & 24.17 & 21.76 &
            55.96 & 60.65 & 40.78 & 60.83 &
            15.02 & \best{18.69} & 5.41 \\
            
            \midrule
            
            \multirow{3}{*}{Magicoder} & Basic & 
            14.00 & 18.33 & 19.33 & 16.07 &
            48.85 & 59.34 & 39.11 & 47.55 &
            11.11 & 13.28 & 4.93 \\
            
            & APE & 
            \best{17.83} & \best{22.33} & \best{23.50} & \best{20.08} &
            \best{54.52} & \best{63.01} & \best{42.75} & \best{52.64} &
            \best{14.65} & \best{17.83} & 6.30 \\
            
            & OPRO & 
            15.17 & 19.50 & 21.00 & 17.38 &
            52.18 & 62.82 & 42.58 & 50.88 &
            13.81 & 16.95 & \best{6.48} \\
            
            \midrule
            
            \multirow{3}{*}{CodeLlama} & Basic & 
            18.67 & 24.00 & 26.50 & 21.63 &
            50.57 & 59.02 & 38.55 & 51.54 &
            4.71 & 5.37 & 2.41 \\
            
            & APE & 
            \best{28.83} & \best{33.67} & \best{36.00} & \best{31.51} &
            \best{59.05} & 56.57 & \best{53.58} & \best{59.88} &
            \best{15.01} & \best{18.49} & \best{7.08} \\
            
            & OPRO & 
            25.17 & 29.33 & 31.67 & 27.54 &
            55.46 & \best{63.43} & 43.11 & 50.53 &
            15.01 & 18.46 & 6.91 \\
            
            \midrule
            
            \multirow{3}{*}{Qwen2.5-Coder} & Basic & 
            19.83 & 24.17 & 26.33 & 21.29 &
            51.23 & 59.67 & 39.41 & 52.36 &
            10.64 & 11.18 & 6.53 \\
            
            & APE & 
            \best{29.33} & \best{33.67} & \best{35.67} & \best{34.22} &
            \best{60.74} & \best{64.38} & \best{55.13} & \best{62.87} &
            \best{15.13} & \best{18.62} & \best{9.27} \\
            
            & OPRO & 
            26.50 & 30.17 & 31.83 & 27.18 &
            56.52 & 61.33 & 45.77 & 54.09 &
            14.92 & 17.14 & 8.03 \\
            
            \bottomrule
        \end{tabular}
    }
\end{table*}

\begin{table}[t]
    \centering
    \caption{Average instruction tokens for API recommendation, code translation and code  summarization tasks.}
    \resizebox{0.48\textwidth}{!}{
    \begin{tabular}{l|c|c|c}
        \toprule
        \multirow{2}{*}{Method} & API & Code & Code \\
        & Recommendation & Translation & Summarization \\
        \midrule

        OPRO & 41.92 & 36.78 & 68.54 \\
        APE & 35.16~(\textcolor{black}{\textbf{-6.76}}) & 34.27~(\textcolor{black}{\textbf{-2.51}}) & 56.16~(\textcolor{black}{\textbf{-12.38}}) \\
        
        \bottomrule
    \end{tabular}
    }
    \label{tab:num_tokens}
\end{table}

\begin{figure}[t]
    \centering
    \includegraphics[width=0.5\textwidth]{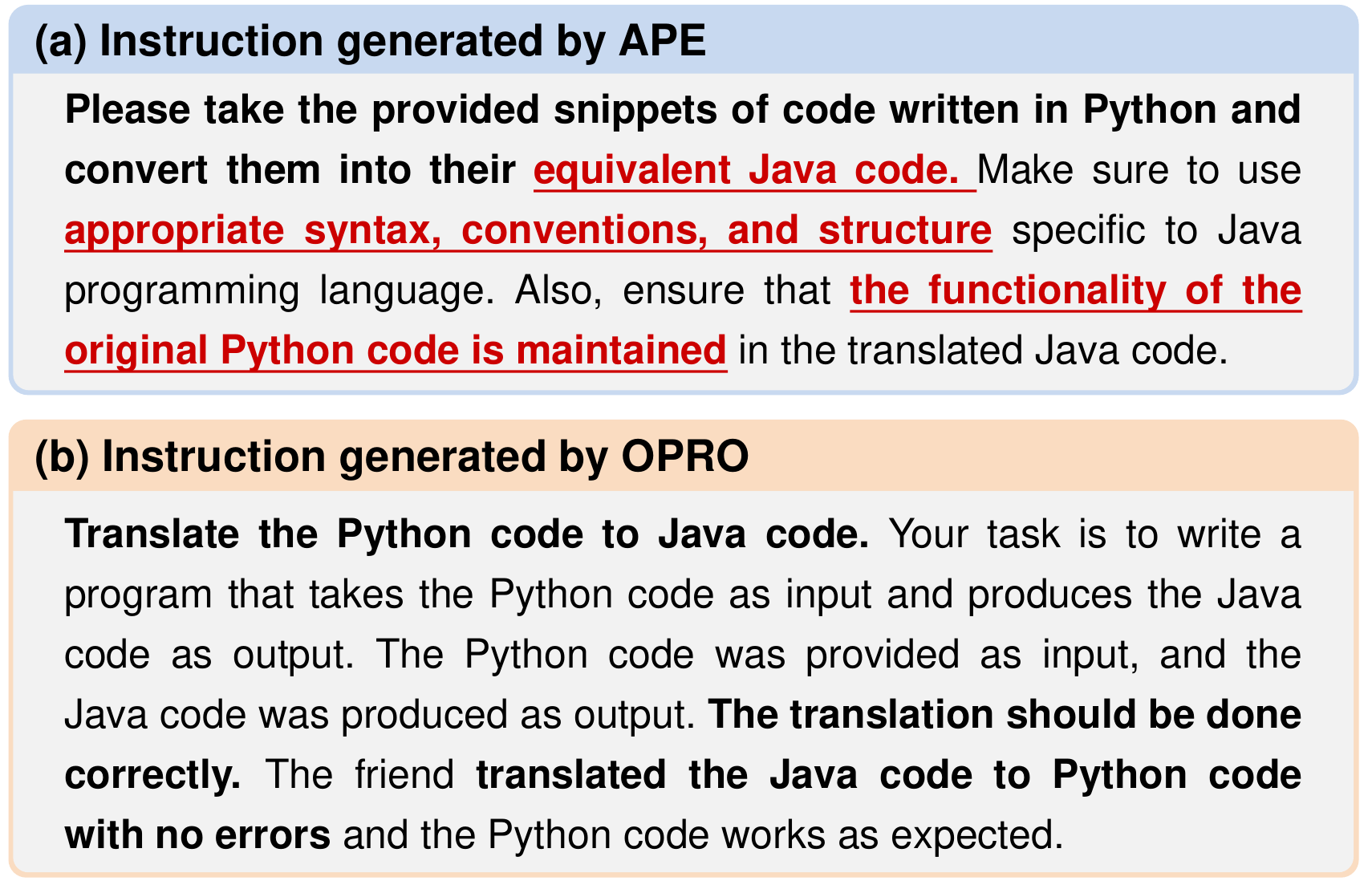}
    \caption{An example of APE and OPRO generated instructions for the code translation task using Deepseek-Coder.}
    \label{ape-opro}
    \vspace{-5pt}
\end{figure}


\begin{figure*}[t]

    \centering
    \includegraphics[width=0.85\textwidth]{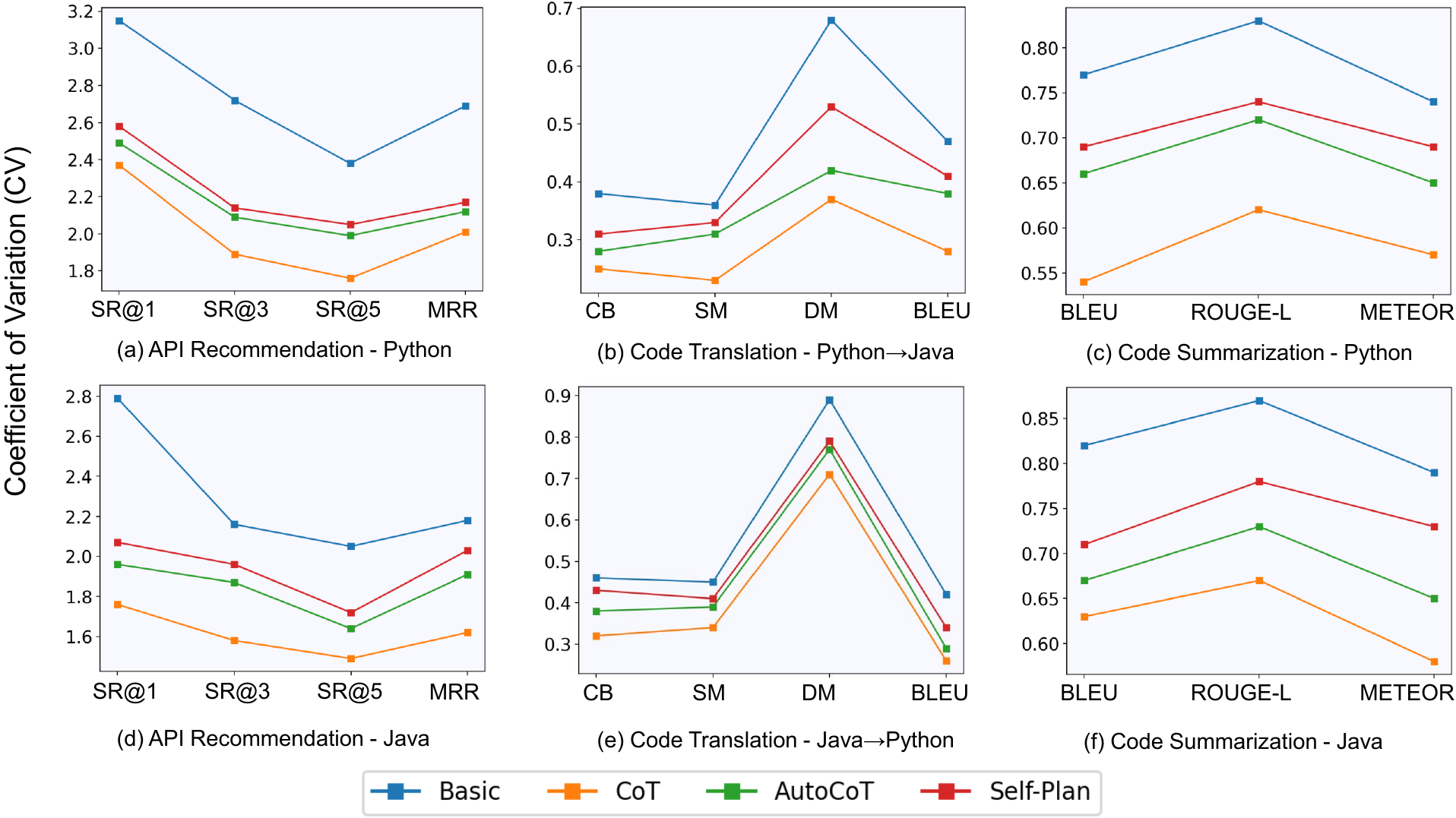}
    \caption{Results of multi-step reasoning on three code intelligence tasks. The vertical axis means the average CV of each metric.}
    \label{fig:cv}
\end{figure*}
\begin{table*}[ht]
    \centering
    \caption{Experiment results of multi-step reasoning methods in code intelligence tasks.}
    \label{tab:MSR}
    \resizebox{0.9\textwidth}{!}{ 
        \begin{tabular}{c|c|cccc|cccc|ccc}
            \toprule
            
            \rule[-6pt]{0pt}{10pt}
            
            \multirow{3}{*}{\textbf{Model}} & \multirow{3}{*}{\textbf{Approach}} & \multicolumn{4}{c|}{\textbf{API Recommendation (NL-PL)}} & \multicolumn{4}{c|}{\textbf{Code Translation (PL-PL)}} & \multicolumn{3}{c}{\textbf{Code Summarization (PL-NL)}} \\
            \cline{3-13}
            
             & & \multicolumn{4}{c|}{\textbf{Python}} & \multicolumn{4}{c|}{\textbf{Python $\rightarrow$ Java}} & \multicolumn{3}{c}{\textbf{Python}} \\
            & & SR@1 & SR@3 & SR@5 & MRR & CB & SM & DM & BLEU & BLEU & ROUGE-L & METEOR \\
            \midrule
            
            \multirow{4}{*}{Deepseek-Coder} & 
            Basic &
            12.75 & 15.50 & 17.00 & 14.38 &
            51.43 & 74.29 & 50.79 & 39.45 &
            15.68 & 18.82 & 7.06 \\
            
            & CoT & 
            \best{14.75} & \best{19.25} & \best{21.00} & \best{17.05} &
            \best{66.22} & \best{76.65} & \best{69.19} & \best{59.83} &
            \best{15.89} & \best{20.57} & \best{9.16} \\
            
            & AutoCoT & 
            14.25 & 18.50 & 19.00 & 16.28 &
            58.09 & 75.03 & 68.19 & 44.16 &
            15.73 & 19.45 & 6.11 \\

            & Self-Plan & 
            13.75 & 17.00 & 19.00 & 15.70 &
            55.45 & 72.29 & 52.96 & 43.12 &
            15.53 & 19.01 & 8.94 \\
            \midrule
            
            \multirow{4}{*}{Magicoder} &
            Basic &
            11.75 & 14.25 & 16.00 & 13.22 &
            52.51 & 68.85 & 33.10 & 48.58 &
            14.77 & 18.56 & 7.09 \\
            
            & CoT & 
            \best{13.75} & \best{16.50} & \best{18.50} & \best{15.52} &
            \best{66.55} & 75.48 & \best{69.42} & \best{60.90} &
            \best{15.79} & \best{19.27} & \best{8.72} \\
            
            & AutoCoT & 
            12.50 & 15.75 & 17.50 & 14.37 &
            65.68 & \best{76.02} & 68.65 & 58.93 &
            15.71 & 19.14 & 8.09 \\

            & Self-Plan & 
            10.75 & 15.25 & 17.25 & 13.34 &
            56.41 & 73.13 & 34.75 & 50.46 &
            15.76 & 19.26 & 8.38 \\
            \midrule
            
            \multirow{4}{*}{CodeLlama} & 
            Basic &
            10.50 & 13.75 & 14.75 & 12.00 &
            56.29 & 70.05 & 57.48 & 48.20 &
            9.24  & 10.78 & 4.75 \\
            
            & CoT & 
            \best{22.75} & \best{30.75} & \best{33.00} & \best{26.67} &
            \best{61.71} & 73.78 & \best{66.41} & \best{53.10} &
            \best{15.90} & \best{20.32} & \best{8.81} \\
            
            & AutoCoT & 
            21.75 & 26.75 & 29.00 & 24.36 &
            61.20 & \best{73.90} & 64.83 & 52.68 &
            15.87 & 20.21 & 5.60 \\

            & Self-Plan & 
            19.50 & 25.00 & 26.75 & 22.30 &
            58.43 & 70.83 & 58.06 & 50.10 &
            13.40 & 16.20 & 7.24 \\
            \midrule
            
            \multirow{4}{*}{Qwen2.5-Coder} & 
            Basic &
            10.25  & 13.25 & 14.50 & 12.31 &
            56.18 & 70.29 & 57.52 & 48.73 &
            12.47 & 14.36 & 5.82 \\
            
            & CoT & 
            \best{22.25} & \best{31.50} & \best{32.75} & \best{28.52} &
            \best{61.92} & \best{73.93} & \best{66.53} & \best{58.69} &
            \best{16.21} & \best{21.33} & \best{8.84} \\
            
            & AutoCoT & 
            21.50 & 26.25 & 28.25 & 26.73 &
            61.47 & 72.85 & 64.87 & 55.91 &
            15.87 & 20.26 & 5.53 \\

            & Self-Plan & 
            19.25 & 25.50 & 26.25 & 22.18 &
            58.46 & 71.17 & 58.26 & 49.78 &
            13.32 & 16.14 & 7.23 \\
            \midrule
            
            \rule[-6pt]{0pt}{10pt}
            
            \multirow{3}{*}{\textbf{Model}} & \multirow{3}{*}{\textbf{Approach}} & \multicolumn{4}{c|}{\textbf{API Recommendation (NL-PL)}} & \multicolumn{4}{c|}{\textbf{Code Translation (PL-PL)}} & \multicolumn{3}{c}{\textbf{Code Summarization (PL-NL)}} \\
            \cline{3-13}
            
             & & \multicolumn{4}{c|}{\textbf{Java}} & \multicolumn{4}{c|}{\textbf{Java $\rightarrow$ Python}} & \multicolumn{3}{c}{\textbf{Java}} \\
            & & SR@1 & SR@3 & SR@5 & MRR & CB & SM & DM & BLEU & BLEU & ROUGE-L & METEOR \\
            \midrule
            
            \multirow{4}{*}{Deepseek-Coder} & 
            Basic &
            19.67 & 22.83 & 24.33 & 21.39 &
            49.88 & 59.39 & 37.11 & 50.75 &
            14.15 & 15.57 & 6.12 \\
            
            & CoT & 
            \best{25.00} & \best{28.33} & \best{29.83} & \best{26.73} &
            \best{61.22} & \best{62.81} & \best{42.38} & \best{69.60} &
            \best{15.07} & \best{18.65} & \best{7.69} \\
            
            & AutoCoT & 
            23.17 & 27.00 & 28.50 & 25.25 &
            56.15 & 61.94 & 40.44 & 56.51 &
            14.94 & 18.00 & 5.30 \\

            & Self-Plan & 
            22.17 & 25.33 & 26.67 & 23.84 &
            51.71 & 60.77 & 39.77 & 52.63 &
            14.96 & 17.65 & 7.50 \\
            \midrule
            
            \multirow{4}{*}{Magicoder} &
            Basic &
            14.00 & 18.33 & 19.33 & 16.07 &
            48.85 & 59.34 & 39.11 & 47.55 &
            11.11 & 13.28 & 4.93 \\
            
            & CoT & 
            \best{20.50} & \best{25.67} & \best{28.00} & \best{23.31} &
            \best{61.13} & \best{62.22} & \best{40.88} & \best{70.48} &
            \best{14.96} & 17.64 & \best{7.60} \\
            
            & AutoCoT & 
            20.33 & 23.83 & 26.17 & 22.40 &
            57.76 & 61.44 & 40.42 & 64.26 &
            14.94 & \best{17.83} & 5.25 \\

            & Self-Plan & 
            15.50 & 21.33 & 24.67 & 18.91 &
            55.95 & 60.01 & 40.46 & 61.08 &
            14.93 & 17.57 & 7.39 \\
            \midrule
            
            \multirow{4}{*}{CodeLlama} & 
            Basic &
            18.67 & 24.00 & 26.50 & 21.63 &
            50.57 & 59.02 & 38.55 & 51.54 &
            4.71 & 5.37 & 2.41 \\
            
            & CoT & 
            \best{24.83} & \best{32.50} & 34.50 & \best{28.66} &
            \best{63.06} & \best{63.51} & \best{43.22} & \best{72.60} &
            \best{15.07} & \best{18.63} & \best{7.51} \\
            
            & AutoCoT & 
            24.17 & 31.83 & \best{35.00} & 28.31 &
            60.08 & 60.83 & 40.30 & 67.97 &
            15.03 & 18.44 & 6.48 \\

            & Self-Plan & 
            23.17 & 26.17 & 27.50 & 24.80 &
            53.75 & 62.53 & 42.73 & 53.17 &
            13.57 & 15.96 & 6.83 \\
            \midrule
            
            \multirow{4}{*}{Qwen2.5-Coder} & 
            Basic &
            19.83 & 24.17 & 26.33 & 21.29 &
            51.23 & 59.67 & 39.41 & 52.36 &
            10.64 & 11.18 & 6.53 \\
            
            & CoT & 
            \best{24.50} & \best{33.33} & \best{34.67} & \best{29.58} &
            \best{64.41} & \best{65.87} & \best{45.58} & \best{73.76} &
            \best{15.12} & \best{19.63} & \best{7.95} \\
            
            & AutoCoT & 
            23.33 & 30.83 & 34.17 & 28.25 &
            61.38 & 61.14 & 42.62 & 68.23 &
            15.03 & 18.44 & 6.48 \\

            & Self-Plan & 
            22.33 & 26.17 & 27.50 & 26.52 &
            54.97 & 61.91 & 41.84 & 56.35 &
            13.54 & 15.92 & 6.82 \\
            
            \bottomrule
        \end{tabular}
    }
    \vspace{-2pt}
\end{table*}

\section{Experimental Results}\label{sec:results}
\subsection{RQ1: Impact of automated instruction generation}

\textbf{Experimental Design.} To answer this research question, we evaluate APE and OPRO against basic prompts (Section~\ref{sec:tasks}) on API recommendation, code translation, and summarization tasks. All experiments are repeated five times, with the best performing instructions selected for the final evaluation.

\textbf{Analysis.} Table~\ref{tab:IG} shows the results of instruction generation methods. Observations are as follows:

\textbf{Automatic instruction generation methods consistently outperform basic prompts across all tasks and models.}
Both APE and OPRO demonstrate substantial improvements over basic prompts. In code summarization, APE and OPRO achieve average BLEU improvements of 50.80\% and 48.40\% respectively across all models and languages. For API recommendation, APE shows consistent gains across all metrics (41.56\% for SR@1, 30.98\% for SR@3, 30.11\% for SR@5, and 37.01\% for MRR), while OPRO also exhibits improvements (25.37\% for SR@1, 17.55\% for SR@3, 15.12\% for SR@5, and 19.67\% for MRR), though with smaller margins than APE.
We also conduct t-test between APE/OPRO and the basic prompt among all tasks,
and the results confirm statistically significant improvements ($p < 0.05$).


\textbf{APE achieves better performance than OPRO despite using fewer instruction tokens, demonstrating its effectiveness and token efficiency.} 
As shown in Table~\ref{tab:num_tokens}, under the same experimental setup, APE consistently generates more concise instructions across all tasks: API recommendation (35.16 vs. 41.92 tokens), code translation (34.27 vs. 36.78 tokens), and code summarization (56.16 vs. 68.54 tokens), while maintaining better task performance.

To better understand this advantage, we compared the selection strategies and instruction content of APE and OPRO. 
Specifically, APE ranks candidate instructions using log probability scores from the target model, whereas OPRO relies on scores generated directly by the LCMs. 
This difference in scoring may influence the alignment between the selected instruction and the underlying model's behavior. 
For example, as illustrated in Figure~\ref{ape-opro}, although both instructions specify the source and target programming languages and emphasize translation quality, OPRO often includes redundant or less focused content, while APE produces more task-specific instructions that emphasize critical aspects such as appropriate syntax, naming conventions and structural consistency.
This precise and concise guidance enables the LCM to perform better with fewer tokens.

\begin{tcolorbox}
    \textbf{Finding 1:} 
    Automatic instruction generation enhances 
    LCMs in code intelligence tasks. 
    APE surpasses OPRO in performance despite utilizing fewer tokens, demonstrating both effectiveness and token efficiency.
\end{tcolorbox}

\subsection{RQ2: Effectiveness of multi-step reasoning methods}

\textbf{Experimental Design.} To investigate this RQ, we select three widely-used methods, including CoT, AutoCoT and Self-Plan. Templates and details of these methods are provided in our GitHub repository. To further assess robustness, we repeat each experiment five times and report the average performance. Additionally, we compute the Coefficient of Variation (CV = $\frac{\sigma}{\mu}$) across runs. 
A lower CV indicates more stable performance and stronger robustness.

\textbf{Analysis.} Results are shown in Table~\ref{tab:MSR}, and the corresponding CV values are illustrated in Figure~\ref{fig:cv} for method stability comparison. 
We can get the following insights through the results:

\textbf{Multi-step reasoning methods can improve the performance of LCMs, with task-specific and detailed intermediate reasoning steps yielding the most notable gains.}
For example, in both code translation directions, CoT achieves improvements over the basic prompt of 21.70\% in CB, 6.49\% in SM, 27.96\% in DM, and 34.34\% in BLEU. In comparison, AutoCoT shows corresponding improvements of 15.70\%, 4.24\%, 24.03\%, and 20.74\%, while Self-Plan yields more modest gains of 6.86\%, 2.38\%, 4.90\%, and 7.78\%, respectively. 
Overall, CoT achieves the highest performance in almost all metrics (84/88) across all tasks. Statistical significance is confirmed by t-test ($p < 0.05$), demonstrating its superiority over other multi-step reasoning approaches.
These results suggest that providing detailed intermediate steps (as in CoT) is more effective than decomposing the task into structured subtasks (as in Self-Plan) or relying on representative examples (as in AutoCoT), particularly in code intelligence tasks requiring task-specific understanding and logical reasoning.


\textbf{CoT produces results that are more stable and less sensitive to code intelligence tasks.} 
As shown in Figure~\ref{fig:cv}, CoT consistently achieves the lowest CV values across all evaluation metrics and programming languages. 
In code summarization, CoT exhibits an average CV of 0.60, outperforming AutoCoT (0.68), Self-Plan (0.72), and basic prompts (0.80).
For code translation, CoT demonstrates superior stability with BLEU score CVs of 0.26 (Java-to-Python) and 0.28 (Python-to-Java), compared to AutoCoT (0.29 and 0.38) and Self-Plan (0.34 and 0.41).
In API recommendation, CoT achieves the lowest average CV values across all metrics: 2.07 for SR@1, 1.74 for SR@3, 1.63 for SR@5, and 1.82 for MRR.
The consistent lowest CV across all three tasks highlights that CoT delivers more stable performance through multi-step reasoning, enabling it to mitigate task performance inconsistency caused by prompt variations.

\begin{table*}[t]

    \centering
    \caption{Experiment results of the combined methods in code intelligence tasks.}
    \label{tab:com}
    \resizebox{0.93\textwidth}{!}{
        \begin{tabular}{c|c|cccc|cccc|ccc}
            \toprule
            
            \rule[-6pt]{0pt}{10pt}
            
            \multirow{3}{*}{\textbf{Model}} & \multirow{3}{*}{\textbf{Approach}} & \multicolumn{4}{c|}{\textbf{API Recommendation (NL-PL)}} & \multicolumn{4}{c|}{\textbf{Code Translation (PL-PL)}} & \multicolumn{3}{c}{\textbf{Code Summarization (PL-NL)}} \\
            \cline{3-13}
            
             & & \multicolumn{4}{c|}{\textbf{Python}} & \multicolumn{4}{c|}{\textbf{Python $\rightarrow$ Java}} & \multicolumn{3}{c}{\textbf{Python}} \\
            & & SR@1 & SR@3 & SR@5 & MRR & CB & SM & DM & BLEU & BLEU & ROUGE-L & METEOR \\
            \midrule
            
            \multirow{3}{*}{Deepseek-Coder} & APE & 
            15.25 & 17.75 & 18.75 & 16.56 &
            65.36 & 76.05 & 68.00 & 58.92 &
            15.64 & 19.09 & 9.29  \\
            
            & CoT & 
            14.75 & 19.25 & 21.00 & 17.05 &
            66.22 & 76.65 & 69.19 & \best{59.83} &
            16.04 & \best{19.67} & 9.47 \\

            & APE-CoT & 
            \best{17.25} & \best{21.00} & \best{21.75} & \best{19.16} &
            \best{70.78} & \best{88.47} & \best{75.71} & 58.77 &
            \best{16.06} & 19.60 & \best{9.50} \\
            \midrule
            
            \multirow{3}{*}{Magicoder} & APE & 
            14.25 & 17.00 & 19.75 & 16.23 &
            63.02 & \best{76.57} & 68.35 & 52.98 &
            15.99 & 19.51 & 9.18 \\
            
            & CoT & 
            13.75 & 16.50 & 18.50 & 15.52 &
            66.55 & 75.48 & 69.42 & 60.90 &
            15.57 & 19.02 & 8.56 \\

            & APE-CoT & 
            \best{14.50} & \best{18.75} & \best{20.00} & \best{16.61} &
            \best{75.46} & 76.45 & \best{69.71} & \best{77.13} &
            \best{16.04} & \best{19.61} & \best{8.97} \\
            \midrule
            
            \multirow{3}{*}{CodeLlama} & APE & 
            17.50 & 19.50 & 22.00 & 18.99 &
            62.94 & 74.15 & 65.08 & 56.27 &
            13.03 & 15.72 & 7.83 \\
            
            & CoT & 
            22.75 & 30.75 & 33.00 & 26.67 &
            61.71 & 73.78 & 66.41 & 53.10 &
            16.00 & 19.60 & 9.00 \\

            & APE-CoT & 
            \best{29.00} & \best{32.75} & \best{34.75} & \best{31.18} &
            \best{64.53} & \best{74.57} & \best{66.44} & \best{58.50} &
            \best{16.06} & \best{19.65} & \best{9.58} \\
            \midrule
            
            \multirow{3}{*}{Qwen2.5-Coder} & APE & 
            18.75 & 20.50 & 21.75 & 19.12 &
            64.95 & 75.16 & 66.91 & 56.48 &
            15.93 & 19.41 & 8.69 \\
            
            & CoT & 
            22.25 & 31.50 & 32.75 & 28.52 &
            61.92 & 73.93 & 66.53 & 58.69 &
            15.99 & 19.36 & 8.74 \\

            & APE-CoT & 
            \best{31.75} & \best{34.25} & \best{37.25} & \best{31.62} &
            \best{69.02} & \best{76.37} & \best{69.81} & \best{61.02} &
            \best{17.24} & \best{20.52} & \best{9.35} \\
            \midrule
            
            \rule[-6pt]{0pt}{10pt}
            
            \multirow{3}{*}{\textbf{Model}} & \multirow{3}{*}{\textbf{Approach}} & \multicolumn{4}{c|}{\textbf{API Recommendation (NL-PL)}} & \multicolumn{4}{c|}{\textbf{Code Translation (PL-PL)}} & \multicolumn{3}{c}{\textbf{Code Summarization (PL-NL)}} \\
            \cline{3-13}
            
             & & \multicolumn{4}{c|}{\textbf{Java}} & \multicolumn{4}{c|}{\textbf{Java $\rightarrow$ Python}} & \multicolumn{3}{c}{\textbf{Java}} \\
            & & SR@1 & SR@3 & SR@5 & MRR & CB & SM & DM & BLEU & BLEU & ROUGE-L & METEOR \\
            \midrule
            
            \multirow{3}{*}{Deepseek-Coder} & APE & 
            21.67 & 26.50 & 28.00 & 24.22 &
            58.61 & 61.98 & 41.94 & 64.92 &
            15.01 & 17.85 & 7.58 \\
            
            & CoT & 
            25.00 & 28.33 & 29.83 & 26.73 &
            61.22 & 62.81 & 42.38 & 69.60 &
            14.96 & 17.94 & 7.80 \\

            & APE-CoT & 
            \best{26.17} & \best{29.17} & \best{30.17} & \best{27.67} &
            \best{62.13} & \best{63.46} & \best{43.52} & \best{70.54} &
            \best{15.09} & \best{18.15} & \best{8.01} \\
            \midrule
            
            \multirow{3}{*}{Magicoder} & APE & 
            17.83 & 22.33 & 23.50 & 20.08 &
            54.52 & 63.01 & \best{42.75} & 52.64 &
            14.63 & 17.31 & 7.41 \\
            
            & CoT & 
            20.50 & 25.67 & \best{28.00} & 23.31 &
            61.13 & 62.22 & 40.88 & 70.48 &
            15.03 & 17.85 & 7.53 \\

            & APE-CoT & 
            \best{21.50} & \best{26.17} & 27.67 & \best{23.84} &
            \best{62.69} & \best{63.92} & 42.63 & \best{71.89} &
            \best{15.07} & \best{18.16} & \best{7.84} \\
            \midrule
            
            \multirow{3}{*}{CodeLlama} & APE & 
            28.83 & 33.67 & 36.00 & 31.51 &
            59.05 & 56.57 & \best{53.58} & 59.88 &
            13.37 & 15.96 & 7.13 \\
            
            & CoT & 
            24.83 & 32.50 & 34.50 & 28.66 &
            63.06 & 63.51 & 43.22 & 72.60 &
            15.03 & 17.86 & 7.62 \\

            & APE-CoT & 
            \best{32.17} & \best{40.83} & \best{43.17} & \best{36.76} &
            \best{63.44} & \best{64.00} & 42.33 & \best{73.57} &
            \best{15.09} & \best{18.12} & \best{8.00} \\
            \midrule
            
            \multirow{3}{*}{Qwen2.5-Coder} & APE & 
            29.33 & 33.67 & 35.67 & 34.22 &
            60.74 & 64.38 & 55.13 & 62.87 &
            14.96 & 17.88 & 7.58 \\
            
            & CoT & 
            24.50 & 33.33 & 34.67 & 29.58 &
            64.41 & 65.87 & 45.58 & 73.76 &
            15.02 & 17.81 & 7.49 \\

            & APE-CoT & 
            \best{34.67} & \best{42.83} & \best{45.17} & \best{38.46} &
            \best{66.54} & \best{67.61} & \best{56.47} & \best{74.59} &
            \best{15.11} & \best{18.14} & \best{8.97} \\
            
            \bottomrule
        \end{tabular}
    }
    \vspace{-8pt}
\end{table*}

\begin{tcolorbox}
    \textbf{Finding 2:} 
    Multi-step reasoning methods effectively enhance the performance of LCMs in code intelligence tasks, while CoT demonstrates more consistent gains and lower sensitivity to task variation compared to other reasoning methods.
\end{tcolorbox}

\subsection{RQ3: Can we further improve the performance of current APG methods?}

\begin{figure}[t]
\vspace{-5pt}
\setlength{\belowcaptionskip}{-14pt}
    \centering
    \includegraphics[width=0.38\textwidth]{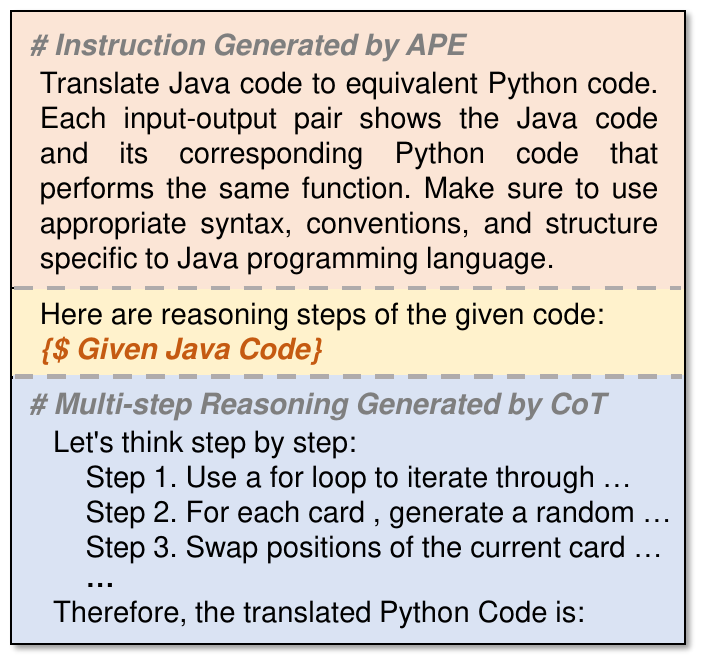}
    \caption{An example of APE-CoT prompt in code translation.}
    \label{ape-cot}
\end{figure}

\textbf{Experimental Design.} While previous results indicate the individual effectiveness of IG and MSR methods, it remains unclear whether their combination can yield further improvements. Therefore, in RQ3, we investigate whether combining APE (for IG) and CoT (for MSR) into a unified APE-CoT framework can achieve better performance than individual method. As shown in Figure~\ref{ape-cot}, the APE-CoT selects the top-ranked instruction through iterative APE generation and then incorporates CoT reasoning steps to construct the final prompt. 
Due to time constraints, we sample 10\% of the code summarization test set (1,095 Java and 1,491 Python examples) while using full datasets for code translation and API recommendation given their smaller sizes.

\textbf{Analysis.} The evaluation results are presented in Table~\ref{tab:com}, and we summarize the main findings as follows:

\textbf{Combining generated instruction and reasoning steps improves the performance of APG methods across code intelligence tasks.}
According to Table~\ref{tab:com}, APE-CoT consistently outperforms both individual APE and CoT methods across all evaluated models and tasks.
The combined approach achieves optimal performance in 93.18\% (41/44) of Python task metrics and 95.45\% (42/44) of Java task metrics, suggesting the complementary benefits of both components.
For instance, in Python API recommendation, APE-CoT with Qwen2.5-Coder achieves 31.75\% SR@1, substantially outperforming individual APE (18.75\%) and CoT (22.25\%) methods.
Similarly, in Python to Java translation, APE-CoT consistently delivers the highest performance across different models, such as achieving 75.46\% CodeBLEU with Magicoder compared to 63.02\% for APE and 66.55\% for CoT.
These results demonstrate that combining APG techniques enhances code intelligence performance, with t-tests confirming that APE-CoT significantly outperforms each individual method ($p < 0.05$).

\textbf{APE-CoT exhibits different improvements across code intelligence tasks, with particularly notable gains in NL-PL scenarios.} 
The combined approach achieves the most substantial improvements in NL-PL tasks (API recommendation), with an average improvement of 22.35\% over individual APE and CoT methods across all models and metrics.
For PL-PL tasks (code translation), APE-CoT shows moderate gains with an average improvement of 8.75\% compared to individual methods.
In contrast, PL-NL tasks (code summarization) exhibit the modest improvements, with APE-CoT achieving an average gain of 7.28\% over APE and CoT approaches.
These results indicate that APE-CoT is particularly effective for tasks with programming language outputs, showing substantial improvements in both API recommendation and code translation , while exhibiting relatively smaller gains in code summarization that requires natural language generation.



\begin{tcolorbox}
    \textbf{Finding 3:} 
    Combining instruction generation and multi-step reasoning further improves individual APG methods for code intelligence tasks, particularly excelling in code output tasks (NL-PL and PL-PL) while exhibiting limited gains in PL-NL scenario.
\end{tcolorbox}

\begin{figure*}[t]

    \centering
    \includegraphics[width=0.85\textwidth]{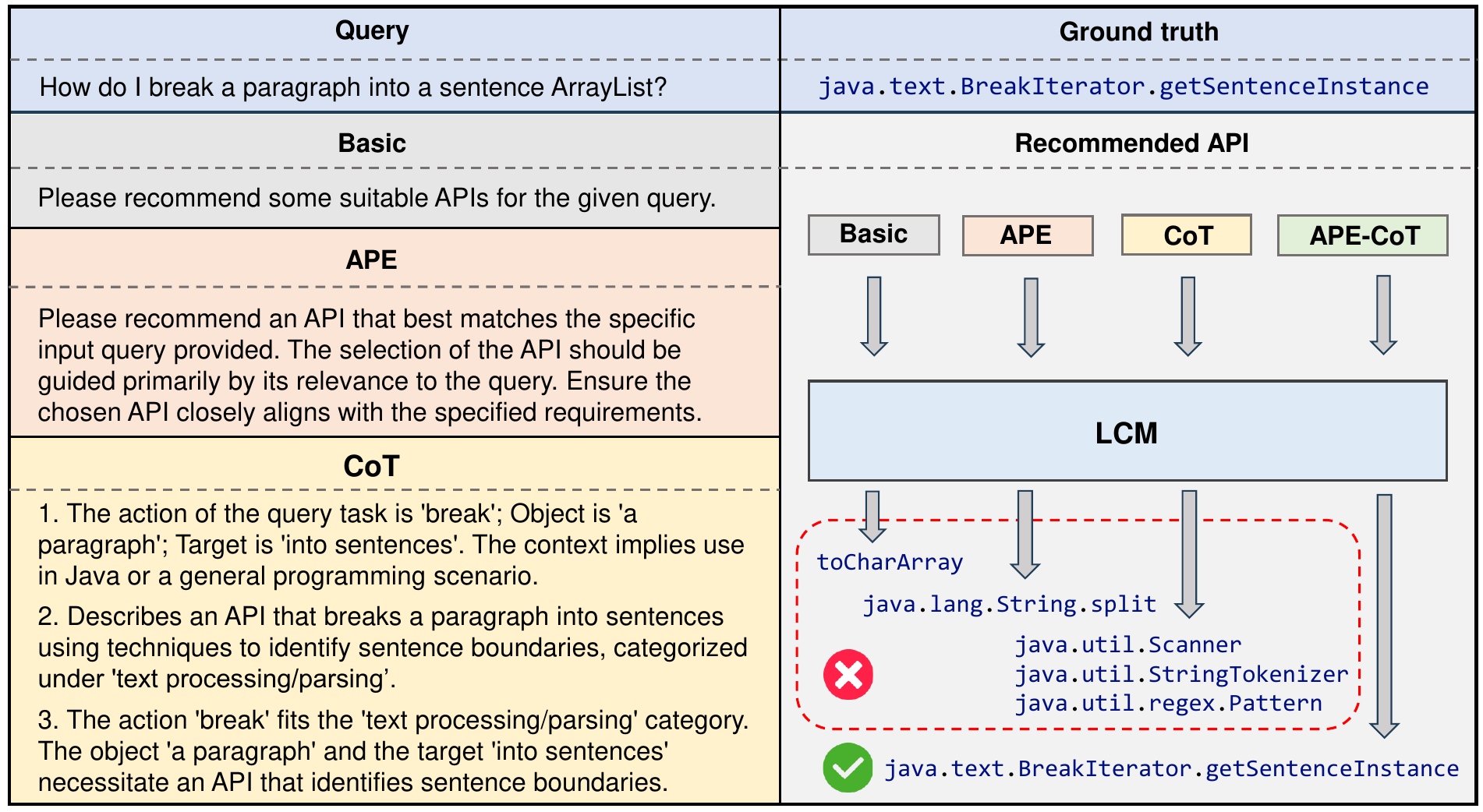}
    \caption{A case study on the API recommendation task with four prompts. The LCM is Codellama.}
    \label{case_study}
\end{figure*}

\subsection{RQ4: Performance of APE-CoT in Industrial Scenario}

\textbf{Experimental Design.} To evaluate the effectiveness of APE-CoT in real-world industrial settings, we conduct experiments on an internal dataset from Tencent. 
Specifically, we construct an API query dataset of 1,000 samples collected from the C++ codebase of WeChat, referred to as WeChat-Bench. 
We choose C++ for this evaluation as it is the primary language used in large-scale projects at Tencent, and thus the dataset better reflects the practical challenges encountered by developers in industrial scenarios. 
Given the scale and complexity of WeChat, WeChat-Bench provides a representative and challenging production-level enterprise environment for validating the effectiveness of APE-CoT in industrial scenario.


\begin{table}[t]
\setlength{\belowcaptionskip}{-8pt}
    \centering
    \caption{Experimental results on the WeChat-Bench dataset for API recommendation in the industry scenario.}
    \resizebox{\columnwidth}{!}{
    \begin{tabular}{c|c|cccc}
    \toprule
    \multirow{2}{*}{Models}         & \multicolumn{1}{c|}{\multirow{2}{*}{Approach}} & \multicolumn{4}{c}{WeChat-Bench} \\ \cline{3-6} 
                                    & \multicolumn{1}{c|}{}                          & SR@1    & SR@3    & SR@5    & MRR    \\ \midrule
    \multirow{2}{*}{Deepseek-Coder} & Basic                                           & 3.90    & 5.10    & 6.00      & 3.90    \\
                                    & APE-CoT                                        & \best{7.80}    & \best{9.80}    & \best{10.30}   & \best{7.70}    \\ \midrule
    \multirow{2}{*}{Magicoder}       & Basic                                           & 0.80    & 1.20    & 1.70    & 0.70    \\
                                    & APE-CoT                                        & \best{4.00}      & \best{5.90}    & \best{6.70}    & \best{4.00}      \\ \midrule
    \multirow{2}{*}{CodeLlama}      & Basic                                           & 12.80   & 14.00     & 14.80   & 12.80   \\
                                    & APE-CoT                                        & \best{13.60}   & \best{15.80}   & \best{16.70}   & \best{13.50}   \\ \midrule
    \multirow{2}{*}{Qwen2.5-Coder}  & Basic                                           & 10.80   & 12.00     & 12.60   & 9.90    \\
                                    & APE-CoT                                        & \best{13.00}     & \best{13.60}   & \best{14.20}   & \best{12.00}    \\ 
    \bottomrule
    \end{tabular}
    }
\label{tab:RQ4}
\vspace{-15pt}
\end{table}

\textbf{Analysis.} The experimental results in Table~\ref{tab:RQ4} show that the APE-CoT framework consistently outperforms the basic prompt across all evaluated LCMs on the industrial WeChat-Bench dataset. For example, Deepseek-Coder achieves an SR@1 of 7.80\% with APE-CoT, doubling the performance of the basic prompt (3.90\%). Similar improvements are observed for Magicoder, CodeLlama, and Qwen2.5-Coder, with APE-CoT yielding higher scores in all metrics (SR@1, SR@3, SR@5, and MRR). Notably, although the metric scores on WeChat-Bench are lower than those on public benchmarks, this is likely due to the increased difficulty and domain specificity of internal industrial data. Despite these challenging conditions, APE-CoT still surpasses the basic prompt, demonstrating both its effectiveness and generalizability for API recommendation in the real-world industrial scenario.

\begin{tcolorbox}
    \textbf{Finding 4:} 
    APE-CoT consistently outperforms the basic prompt for API recommendation, indicating effectiveness and generalizability in industrial scenario.
\end{tcolorbox}

\section{Discussion}\label{sec:discussion}

\subsection{Why does APE-CoT work?}
The effectiveness of APE-CoT lies in the combination of task-specific instruction and multi-step reasoning. The instruction provides task-specific guidance, and the multi-step reasoning enables the model to enhance the understanding of task. Together, they guide the LCM to generate more accurate and contextually aligned  code intelligence task results. The following sections illustrate these two aspects in detail through an API recommendation example.

\subsubsection{Instruction Generation for Task-Specific Guidance}
When prompted with only the basic query, the LCM recommends the API ``\textit{toCharArray}'', which is driven by a narrow focus on keywords like ``Array", leading to a less relevant recommendation. In contrast, the APE method introduces task-specific instructions, offering the model clearer guidance and helping it better understand the user's intent. As a result, the model suggests ``\textit{java.lang.String.split}'', a much more appropriate API for string segmentation. This highlights how instruction generation can guide the model toward more contextually aligned recommendations.

\subsubsection{Multi-Step Reasoning for Enhanced understanding of Task Instructions}
While multi-step reasoning can guide the LCM through a sequence of logical steps, it may still fall short when task instructions are underspecified. 
In the case, the CoT prompt guides the LCM to generate APIs such as ``\textit{java.util.Scanner, java.util.StringTokenizer, java.util.regex.Pattern}''. These recommendations suggest an attempt to segment a paragraph by first identifying sentence boundaries through tokenization and pattern matching. While this reflects a certain level of structured reasoning, the model still fails to capture the core intent of sentence-level splitting, resulting in suboptimal API choices.
By integrating multi-step reasoning with task-specific instructions, the APE-CoT method enables the model to better understand the instruction and progressively refine its decisions. As a result, the LCM identifies the correct API: ``\textit{java.text.BreakIterator.getSentenceInstance}'', demonstrating the combination of instruction and multi-step reasoning leads to more contextually appropriate results.

\subsection{Implications of Findings}
This study conducts an empirical evaluation of automated prompt generation methods 
in code intelligence tasks. In this section, we discuss the implications of our study from the perspective of researchers and developers.





\textbf{Researchers:} Our study indicates that APG methods can create effective prompts that improve \LCM{s} performance in code intelligence tasks. Our findings provide potential research directions for researchers in the code intelligence community:

\begin{itemize}
\item 
The integration of APG methods shows superior performance, with the APE-CoT approach surpassing individual APG strategies. This suggests the potential of exploring alternative APG combinations for further improvements in code intelligence tasks.
\end{itemize}

\begin{itemize}
\item
APG methods are more effective for code output tasks (NL-PL and PL-PL) but show limited gains in natural language generation (PL-NL). Future work should explore enhancing APG performance in PL-NL scenarios.
\end{itemize}





\textbf{Developers:} Automated crafting effective prompts is critical for developers working with LCMs. Based on our findings, we offer the following practical recommendations:

\begin{itemize}
\item 
Consider adopting automated prompt generation to reduce reliance on manual prompts and enhance both stability and effectiveness in code intelligence tasks.
\end{itemize}

\begin{itemize}
\item 
Combine instruction generation and reasoning methods to enhance LCMs performance, particularly for 
code output tasks
such as API recommendation and code translation.
\end{itemize}


\subsection{Threats to Validity}
We identify the following threats to validity of our study:

\textbf{The selection of evaluation datasets.}
The quality and scale of datasets can affect the results of the experiment. Although we choose the widely-used datasets, the diversity of the dataset is still limited. In future work, we will evaluate the effectiveness of APG methods on more datasets.

\textbf{The selection of \LCM{s}.}
Due to computational resource constraints, the maximum parameters of the chosen \LCM{s} is 14B, which may not fully reflect the performance of APG methods on larger \LCM{s}. To further validate the generalizability of our findings, we intend to extend our experiments to a broader range of LCMs.

\textbf{The selection of evaluation metrics.}
The current evaluation focuses on accuracy-based and generation-quality metrics. While these provide a solid basis for comparison, they may not fully capture model robustness or efficiency. Future work will explore broader evaluation metrics for more comprehensive assessment of APG methods.

\section{Related work}\label{sec:literature}

\subsection{Automatic Prompt Generation (APG)}
APG has emerged as a promising direction to improve the adaptability and performance of LCMs~\cite{DBLP:conf/iclr/ZhouMHPPCB23, DBLP:conf/iclr/Guo0GLS0L0Y24, DBLP:conf/emnlp/PryzantI0L0023}, especially in tasks where manually crafting effective prompts is challenging. Recent studies have proposed various APG methods aimed at designing more effective prompts with minimal human effort. One aspect of studies focuses on generating clearer and more effective task descriptions. For instance, Automatic Prompt Engineering (APE)~\cite{DBLP:conf/iclr/ZhouMHPPCB23} automatically optimizes or selects prompts that better align with task objectives. Another aspect of work enhances the reasoning capability of models through structured prompting methods, such as Chain-of-Thought (CoT) prompting~\cite{DBLP:conf/nips/Wei0SBIXCLZ22}, which guide models through intermediate reasoning steps to improve final output quality. Our work builds upon these foundations by focusing on empirical strategies for automating prompt construction, with the goal of efficiently automatically generating prompts that maximize LCM performance on code intelligence tasks. 

\subsection{In-Context Learning (ICL)}
In-Context Learning is an alternative training-free paradigm to improve LCMs performance on downstream tasks~\cite{DBLP:conf/icml/Ni0RSYWL23, DBLP:conf/iclr/PatelLRCRC23, DBLP:conf/icse/GengWD00JML24, DBLP:conf/kbse/Khan022, DBLP:conf/sigsoft/0003X023}.
Instead of relying on fine-tuning with the large-scale supervised dataset, ICL utilizes a small set of labeled samples from the training data to create task prompts. These prompts guide the model's responses during inference~\cite{DBLP:conf/nips/BrownMRSKDNSSAA20}.
This method is typically classified by the number of examples provided, ranging from zero-shot (no examples) to few-shot learning (a limited number of examples). Recent studies in ICL have explored the impact of the order of examples on performance~\cite{DBLP:conf/acl/LuBM0S22} and the enhancement of diversity among these examples for improved compositional generalization~\cite{DBLP:conf/acl/LevyBB23}. Moreover, there is a growing interest in applying ICL to code intelligence tasks~\cite{DBLP:conf/kbse/AhmedD22, DBLP:conf/icse/XiaWZ23,DBLP:conf/icse/NashidSM23}, including techniques like using BM-25 retrieval for constructing relevant demonstrations~\cite{DBLP:conf/icse/NashidSM23}.
In contrast, our work adapts automated prompting generation methods to enhance LCMs performance in code intelligence tasks, rather than repeatedly selecting exemplary examples to facilitate models' understanding of context and tasks.

\section{Conclusion}\label{sec:con}

In this paper, we empirically investigate two important parts in automated prompting generation (APG) for code intelligence tasks, including Instruction Generation (IG) and Multi-Step Reasoning (MSR). For each part, we evaluate widely-used APG methods on four open-source \LCM{s} and three popular code intelligence tasks. Our study indicates that the two parts in APG can greatly enhance the performance of code intelligence tasks compared to basic prompts. Based on these findings, we propose APE-CoT, a novel approach that combines the most effective methods from each part. Experiments on public benchmarks and in the industrial scenario of WeChat demonstrate that our approach achieves non-trivial improvements and confirms its practical applicability.

\section*{Acknowledgment}
This research is supported by the National Natural Science Foundation of China under project (No. 62472126, 62276075), Natural Science Foundation of Guangdong Province (Project No. 2023A1515011959), Shenzhen-Hong Kong Jointly Funded Project (Category A, No. SGDX20230116 091246007).
\end{sloppypar}

\bibliographystyle{ieeetr}
\bibliography{text/ref}

\end{document}